\documentclass [prd,reprint,eqsecnum,nofootinbib]  {revtex4-1}
\usepackage{amsfonts,amsmath,amscd,mathrsfs,hyperref,dsfont}
\def\SO{\mathsf{SO}}
\def\so{\mathsf{so}}
\def\SU{\mathsf{SU}}
\def\cP{{\cal P}}
\def\cD{{\cal D}}
\def\cC{{\cal C}}
\def\cL{{\cal L}}

\begin{document}
\title{%
Hamiltonian analysis of $\SO(4,1)$ constrained BF theory}
\author{R. Durka}
\email{rdurka@ift.uni.wroc.pl}\affiliation{Institute for Theoretical Physics,
University of Wroc\l{}aw, Pl.\ Maxa Borna 9, Pl--50-204 Wroc\l{}aw, Poland}
\author{J. Kowalski-Glikman}
\email{jkowalskiglikman@ift.uni.wroc.pl}\affiliation{Institute for Theoretical Physics,
University of Wroc\l{}aw, Pl.\ Maxa Borna 9, Pl--50-204 Wroc\l{}aw, Poland}
\date{\today}
\begin{abstract}
In this paper we discuss canonical analysis of $\SO(4,1)$ constrained BF theory. The action of this theory contains  topological terms appended by a term that breaks the gauge symmetry down to the Lorentz subgroup of $\SO(3,1)$. The equations of motion of this theory turn out to be the vacuum Einstein equations. By solving the $B$ field equations one finds that the action of this theory contains not only the standard Einstein-Cartan term, but also the Holst term proportional to the inverse of the Immirzi parameter, as well as a combination of topological invariants. We show that the structure of the constraints of a $\SO(4,1)$ constrained BF theory is exactly that of gravity in Holst formulation. We also briefly discuss quantization of the theory.
\end{abstract}
\maketitle
\section{Introduction}
One of the most remarkable developments in general relativity of the last decades was Ashtekar's discovery that the phase space of gravity can be described with the help of a background independent theory of self-dual $\SU(2)$ connection \cite{Ashtekar:1986yd}. This discovery became a foundation of the research program of Loop Quantum Gravity \cite{Rovelli:2004tv}, \cite{Thiemann:2007zz}. The original Ashtekar's formulation was generalized few years later by Barbero to the case of real connections \cite{Barbero:1994ap}, parametrized by a single real number $\gamma$, called the Immirzi parameter \cite{Immirzi:1996di}. It turns out that this parameter is in fact an additional dimensionless coupling constant of the gravitational action, which takes the symbolic form \cite{Holst:1995pc}
\begin{equation}\label{0.1}
    S^{grav} = \frac1G\int e_\mu^\alpha e_\nu^\beta R_{\rho\sigma}{}_{\gamma\delta}\left( \epsilon_{\alpha\beta}{}^{\gamma\delta}+\frac1{\gamma }\,\delta^{\gamma\delta}_{\alpha\beta}\right)\epsilon^{\mu\nu\rho\sigma} -\frac{\Lambda}{3G} e^4\, ,
\end{equation}
$$
e^4\equiv \epsilon^{\mu\nu\rho\sigma}\, \epsilon_{\alpha\beta\gamma\delta} e_\mu^\alpha e_\nu^\beta e_\rho^\gamma e_\sigma^\delta\, .
$$
In the action above $e^\alpha$ is the tetrad one-form and $R_{\alpha\beta}$ is the curvature two-form of the Lorentz connection $\omega_{\alpha\beta}$, where the Lorentz algebra indices $\alpha,\beta, \ldots$ run from 0 to 3.  Normally the second term, called the Holst term, regardless of {\em not} being a total derivative, does not affect field equations, because its contribution vanishes  on shell (for zero torsion) by virtue of the Bianchi identity. In spite of this, its presence is not completely innocent: it affects canonical structure of the classical theory, and quantum theories for different $\gamma$ lead to different physical predictions (for example the expression for black hole entropy calculated in this framework depends on $\gamma$ \cite{Rovelli:1996dv}).

It has been noticed in \cite{Rezende:2009sv} that from Wilsonian perspective it would be quite unnatural not to append the action (\ref{0.1}) with all possible terms  that are compatible with the field content ($e$ and $\omega$) and (local Lorentz and diffeomorphism) symmetries of the theory. It turns out that there are only three such terms corresponding to three topological invariants (Pontryagin, Euler and Nieh-Yan classes, see (\ref{ny})--(\ref{eu}) below). Again, the presence of these terms does not influence the classical field equation when the constant time slices of the spacetime are compact without boundaries. However, they may play an important role in quantum theory and/or in the case when boundaries are present. In the formulation of \cite{Rezende:2009sv} all these terms come with a priori independent coupling constants and one wonders if it would be possible to find a formulation of the theory so as to organize them in a unified way.

Such a formulation is known for quite some time and is dubbed constrained BF theory. The idea that gravity can be formulated as a constrained topological
BF theory has its roots in works of  MacDowell and Mansouri
\cite{MacDowell:1977jt} and of Plebanski \cite{Plebanski:1977zz}.
The starting point of the present work will be the following action,
proposed and discussed in \cite{Freidel:2005ak} (see also
\cite{Smolin:2003qu}),
$$
\mathcal{S}=\int d^4x\, \epsilon^{\mu\nu\lambda\rho}\, \Big(B_{\mu\nu
IJ}F_{\lambda\rho}^{IJ}-\frac{\beta}{2}B_{\mu\nu
IJ}B^{IJ}_{\lambda\rho}
$$
\begin{equation}\label{1}
-\frac{\alpha}{4}\epsilon_{IJKL4}B^{IJ}_{\mu\nu}B^{KL}_{\lambda\rho}\Big)\, .
\end{equation}
In this action
$$F_{\mu\nu}{}^{IJ} =\partial_\mu A_\nu{}^{IJ}-\partial_\nu A_\mu{}^{IJ} +A_\mu{}^I{}_K\, A_\nu{}^{KJ}  -A_\nu{}^I{}_K\, A_\mu{}^{KJ}$$
is the field strength of the $\SO(4,1)$ (or $\SO(3,2)$)  connection
$A_\mu{}^{IJ}$, while $B_{\mu\nu}{}^{IJ}$ is a two-form field valued
in the algebra of the same gauge group. The capital Latin indices
$I,J,K,\ldots$ are the algebra ones and run from 0 to 4, when the
Lorentz subalgebra of the gauge algebra is labeled by Greek indices
from the beginning of the alphabet $\alpha,\beta,\gamma,\ldots$
running from 0 to 3. We will decompose them into timelike $0$ and
spacelike $a,b,c,\ldots$. Below, in the course of Hamiltonian
analysis we also decompose the spacetime indices $\mu,\nu$ into time
and space denoting the space indices by  letters from the
middle of the Latin alphabet $i,j,k,\ldots$.

As we will show in the next section, the theory defined by the action (\ref{1}) is equivalent to Einstein--Cartan theory with action accompanied with the Holst term and the topological terms described above. The six coupling constants of \cite{Rezende:2009sv} are then replaced by two dimensionless couplings $\alpha$ and $\beta$ of (\ref{1}) and one dimensionful scale $\ell$.  In the Sec. II we will discuss the canonical formulation of this theory, while in Secs. IV and V we will show how these constraints can be simplified and recast into the form proposed by Holst. In the final section we will make some comments concerning perturbative quantization of the theory around Kodama state.

\section{Gravity as a constrained BF theory}

In this section we will recall some properties of the action (\ref{1}). It has been shown in \cite{Freidel:2005ak} that this action
is equivalent to the standard action of Einstein-Cartan gravity. To
see this one first decomposes the connection $A_\mu{}^{IJ}$ into
tetrad and Lorentz connection
\begin{equation}\label{2}
    A_\mu{}^{\alpha 4}=\frac1\ell\, e^\alpha_\mu, \quad A_\mu{}^{\alpha\beta}=\omega_\mu{}^{\alpha\beta}\, ,
\end{equation}
with $\ell$ being a length scale, necessary for dimensional  reasons
since the connection on the left hand side has the dimension of
inverse length, while tetrad is dimensionless\footnote{In our
approach all generators of the gauge algebra are dimensionless.
Alternatively, one can use dimensionful generators
of the translational part of the algebra (as it is usually done when one wants
eventually to make the algebra contraction). Then momentum generators have
canonical dimension of inverse length and $\ell$ shows up in
the algebra as well.} associated with the cosmological constant
$$\frac{1}{\ell^2}= \frac{\Lambda}{3}$$
Then one solves equations of motion for $B$ and substitutes the result back into  the
action. As a result one finds Einstein action appended with a number
of topological invariants.
To find its canonical form one has to
associate  the dimensionless coupling constants $\alpha$ and $\beta$
of (\ref{1}) with the physical ones: Newton's constant $G$, the
cosmological constant $\Lambda$, and the Immirzi parameter $\gamma$:
\begin{equation}\label{3}
    \alpha =
\frac{G\Lambda}{3}\frac{1}{(1+\gamma^2)}, \quad \beta =
\frac{G\Lambda}{3}\frac{\gamma}{(1+\gamma^2)} , \quad \gamma=\frac{\beta}{\alpha}\, .
\end{equation}

Instead of repeating this derivation here, let us show that field
equations resulting from the action (\ref{1}) are the standard
vacuum Einstein equations. The field equations read
\begin{equation}\label{2.1}
   \epsilon^{\mu\nu\rho\sigma}( {\cal D}^A_{\mu}\, B_{\nu\rho})^{IJ}=0\, ,
\end{equation}
\begin{equation}\label{2.2}
   \epsilon^{\mu\nu\rho\sigma}\left( F_{\mu\nu}{}^{IJ} -\beta\, B_{\mu\nu}{}^{IJ}-\frac\alpha2\, \epsilon^{IJKL4}\, B_{\mu\nu}{}_{KL}\right)=0\, .
\end{equation}
In (\ref{2.1}) ${\cal D}^A_{\mu}$ is the covariant  derivative
defined by connection $A$, so that
$$
( {\cal D}^A_{\mu}\, B_{\nu\rho})^{IJ}=\partial_{\mu}\, B_{\nu\rho}{}^{IJ} +A_{\mu}^I{}_K\, B_{\nu\rho}{}^{KJ}+ A_{\mu}^J{}_K\, B_{\nu\rho}{}^{IK}\, .
$$
The theory defined by (\ref{1}) for non-zero $\alpha$  breaks the
original de Sitter $\SO(4,1)$ gauge symmetry down to Lorentz
$\SO(3,1)$. It is, therefore, convenient to decompose the covariant
derivative ${\cal D}^A_{\mu}$ into Lorentz $\so(3,1)$ and
translational parts, and to use the Lorentz covariant derivative
defined by Lorentz connection $\omega$ (\ref{2}), to wit
\begin{equation}\label{2.3}
   ( {\cal D}^A_{\mu}\, B_{\nu\rho})^{\alpha\beta}= ({\cal D}^\omega_{\mu}\, B_{\nu\rho})^{\alpha\beta} - \frac1\ell\, e_{\mu}{}^\alpha\, B_{\nu\rho}{}^{\beta4}+\frac1\ell\, e_{\mu}{}^\beta\, B_{\nu\rho}{}^{\alpha4}\, ,
\end{equation}
\begin{equation}\label{2.4}
   ( {\cal D}^A_{\mu}\, B_{\nu\rho})^{\alpha4}=({\cal D}^\omega_{\mu}\, B_{\nu\rho})^{\alpha4} - \frac1\ell\, e_{\mu}{}_\beta\, B_{\nu\rho}{}^{\alpha\beta}\, ,
\end{equation}
where
\begin{equation}\label{2.5}
     ( {\cal D}^\omega_{\mu}\, B_{\nu\rho})^{\alpha\beta}=\partial_{\mu}\, B_{\nu\rho}{}^{\alpha\beta}+\omega_{\mu}{}^\alpha{}_\gamma\, B_{\nu\rho}{}^{\gamma\beta}+ \omega_{\mu}{}^\beta{}_\gamma\, B_{\nu\rho}{}^{\alpha\gamma}
\end{equation}
with an obvious generalization for another Lorentz tensors.  Using
this decomposition we rewrite the field equations (\ref{2.2}),
(\ref{2.3}) as
\begin{equation}\label{2.6}
   \epsilon^{\mu\nu\rho\sigma}\left( {\cal D}^\omega_{\mu}\, B_{\nu\rho}{}^{\alpha\beta} - \frac1\ell\, e_{\mu}{}^\alpha\, B_{\nu\rho}{}^{\beta4}+\frac1\ell\, e_{\mu}{}^\beta\, B_{\nu\rho}{}^{\alpha4}\right)=0\, ,
\end{equation}
\begin{equation}\label{2.7}
   \epsilon^{\mu\nu\rho\sigma}\left( {\cal D}^\omega_{\mu}\, B_{\nu\rho}{}^{\alpha4} - \frac1\ell\, e_{\mu}{}_\beta\, B_{\nu\rho}{}^{\alpha\beta}\right)=0\, ,
\end{equation}
\begin{equation}\label{2.8}
   F_{\mu\nu}{}^{\alpha\beta} -\beta\, B_{\mu\nu}{}^{\alpha\beta}-\frac\alpha2\, \epsilon^{\alpha\beta\gamma\delta}\, B_{\mu\nu}{}_{\,\gamma\delta}=0\, ,
\end{equation}
\begin{equation}\label{2.9}
   F_{\mu\nu}{}^{\alpha 4} -\beta\, B_{\mu\nu}{}^{\alpha 4}=0\, .
\end{equation}
Notice that the curvature in (\ref{2.8}) is the sum of Riemann tensor of $\omega$ and the cosmological curvature
\begin{equation}\label{2.10}
    F_{\mu\nu}{}^{\alpha\beta}=R_{\mu\nu}{}^{\alpha\beta} -\frac1{\ell^2}\left(e_{\mu}{}^\alpha\,e_{\nu}{}^\beta-e_{\nu}{}^\alpha\,e_{\mu}{}^\beta\right)\, ,
\end{equation}
while that in (\ref{2.9}) is just the torsion
\begin{equation}\label{2.11}
    F_{\mu\nu}{}^{\alpha 4} =\frac{1}{\ell}\big({\cal D}^\omega_{\mu}\, e_{\nu}{}^{\alpha}-{\cal D}^\omega_{\nu}\, e_{\mu}{}^{\alpha}\big)=\frac{1}{\ell}T_{\mu\nu}{}^{\alpha}\, .
\end{equation}
Solving (\ref{2.8}) and  (\ref{2.9}) for $B$ we find
\begin{equation}\label{2.12}
    B_{\mu\nu}{}^{\alpha 4} = \frac1\beta\, F_{\mu\nu}{}^{\alpha 4}, \quad B_{\mu\nu}{}^{\alpha\beta} = \frac{1}{2}M^{\alpha\beta}{}_{\gamma\delta}\, F_{\mu\nu}{}^{\gamma\delta}\, ,
\end{equation}
where
\begin{equation}\label{2.12a}
   M^{\alpha\beta}{}_{\gamma\delta}= \frac{1}{(\alpha^2+\beta^2)}( \beta \delta^{\alpha\beta}_{\gamma\delta}-\alpha \epsilon^{\alpha\beta}{}_{\gamma\delta})\, ,
\end{equation}
with $\displaystyle
 \delta^{\alpha\beta}_{\gamma\delta} \equiv \delta^{\alpha}_{\gamma}\delta^{\beta}_{\delta}-\delta^{\beta}_{\gamma}\delta^{\alpha}_{\delta}\, .
$
The tensor $M$ is  a sum of Lorentz invariant tensors and, therefore, its covariant derivative ${\cal D}^\omega_{\mu}$ vanishes.

Substituting (\ref{2.12}) into (\ref{2.6}) and using Bianchi identity
for Riemann  curvature one can check that the resulting equation
forces torsion $T_{\mu\nu}{}^{\alpha} =\ell F_{\mu\nu}{}^{\alpha 4}$
to vanish\footnote{To prove this one has to assume invertibility of the
tetrad.}. Using this it is easy to see that (\ref{2.7}) is
equivalent to Einstein equations with cosmological constant
$\Lambda=3/\ell^2$. This completes the proof that field equations
following from the action (\ref{1}) reproduce the standard Einstein
equations.

It should be noticed that when the coupling constant $\alpha=0$ the
theory becomes  topological, so that the last term in the action
(\ref{1}) that explicitly breaks the gauge symmetry from the
topological $\SO(4,1)$ down to physical $\SO(3,1)$ carries all the
information about dynamical local degrees of freedom of gravity. As
we will see below this fact is clearly reflected in the structure of
constraints algebra.

\section{Canonical analysis}

In the first step of canonical analysis of the constrained BF theory
defined by (\ref{1})  let us decompose the curvature
$F_{\mu\nu}{}^{IJ}$ into electric and magnetic parts
\begin{equation}\label{4}
    F_{\mu\nu}{}^{IJ} \rightarrow (F_{0i}{}^{IJ}, F_{ij}{}^{IJ})
\end{equation}
with
$$
    F_{0i}{}^{IJ} =\dot A_i{}^{IJ} -\partial_iA_0{}^{IJ}+A_0{}^I{}_K\, A_i{}^{KJ} - A_i{}^I{}_K\, A_0{}^{KJ}$$\begin{equation}\label{5}=\dot A_i{}^{IJ} - \cD_i A_0{}^{IJ}
\end{equation}
where the dot denotes the time derivative,  $\cD_i$ is the
covariant derivative for the connection $A_i{}^{IJ}$, and
\begin{equation}\label{6}
    F_{ij}{}^{IJ} =\partial_i A_j{}^{IJ} +A_i{}^I{}_K\, A_j{}^{KJ} - i \leftrightarrow j\, .
\end{equation}
As usual the zero component of the connection  becomes a Lagrange multiplier for Gauss law. Further we decompose $B$ field into
\begin{equation}\label{7}
    B_{\mu\nu}{}^{IJ} \rightarrow \left(B_{0i}{}^{IJ} \equiv B_{i}{}^{IJ},\, \cP^{i}{}^{IJ}\equiv 2\epsilon^{ijk}\, B_{jk}{}^{IJ}\right)\, .
\end{equation}
As we will see shortly, $\cP^{i}{}^{IJ}$ turn out to be momenta associated with spacial components of gauge field $A$, while the remaining components of $B$  play a role of Lagrange multipliers.

Using these definitions and integrating by parts  we can rewrite the action as follows
\begin{equation}\label{8}
   S=\int dt L\, ,
\end{equation}
\begin{equation}\label{9}
    L=\int d^3x \big(\cP^{i}{}_{IJ} \dot A_i{}^{IJ} + B_{i}{}^{IJ} \Pi^i{}_{IJ} + A_0{}^{IJ}
    \Pi_{IJ}\big)\, .
\end{equation}
It is clear that $B_{i}{}^{IJ}$ and $A_0{}^{IJ}$ are Lagrange
multipliers enforcing the constraints $\Pi^i{}_{IJ}$ and $\Pi_{IJ}$,
which  explicitly read:
$$
   \Pi_{IJ}(x) = \left(\cD_i \cP^i\right)_{IJ}(x)
   $$\begin{equation}\label{10}
   =\Big(\partial_i\cP^i{}_{IJ} + A_i{}_I{}^K\cP^i{}_{KJ}+A_i{}_J{}^K\cP^i{}_{IK}\Big) (x)
\end{equation}
which is the Gauss law for $\SO(4,1)$ invariance (see below), and
\begin{equation}\label{11}
   \Pi^i{}_{IJ}(x) = \left(2\epsilon^{ijk}\, F_{jk}{}_{IJ}-\beta\, \cP^{i}_{IJ} - \frac\alpha2\, \epsilon_{IJKL4}\, \cP^i{}^{KL}\right)(x)
\end{equation}
The Poisson bracket of the theory is
\begin{equation}\label{12}
   \left\{ A_i{}^{IJ}(x), \cP^j{}_{KL}(y)\right\}=\frac12\, \delta(x-y)\, \delta_i^j\,
   \delta_{KL}^{IJ}\, .
\end{equation}
(The factor $1/2$ results from the fact that the canonical momentum
associated with $A$ defined as $\delta L/ \delta \dot A$ is
$2\cP$, not $\cP$.) The Lagrangian (\ref{9}) contains just the
standard ($p\dot q$) kinetic term appended with a combination of
constraints, reflecting the manifestation of diffeomorphism
invariance of the action (\ref{1}) that we have started with. It is
worth noticing that prior to taking care of the constraints the
dimension of phase space of the system is $2 \times 3 \times 10 =60$
at each space point. As we will see the dimension of the physical
phase space is going to be $4$, as it should be.

The Poisson brackets of the constraints can be straightforwardly computed and read\\[10pt]
$
   \{\Pi_{IJ}(x), \Pi_{KL}(y)\}= \delta(x-y)\, \Big(\eta_{IL}\Pi_{JK}(x) $
\begin{equation}\label{13}  -\eta_{JL}\Pi_{IK}(x)-\eta_{IK}\Pi_{JL}(x) +\eta_{JK}\Pi_{IL}(x)\Big)\approx0
\end{equation}
which means that $\Pi_{IJ}$ form a representation of the gauge group $SO(4,1)$ of the unconstrained theory ($\alpha=0$), as expected.
Further\\[10pt]
$\{\Pi^i{}_{IJ}(x), \Pi^j{}_{KL}(y)\}=2\alpha\epsilon^{ijk}\, \delta(x-y)
\,\Big(\epsilon_{KLIP4}A_k{}^P{}_J(x)$
\begin{equation}\label{14} - \epsilon_{KLJP4}A_k{}^P{}_I(x) +\epsilon_{IJKP4}A_k{}^P{}_L(x)-\epsilon_{IJLP4}A_k{}^P{}_K(x)\Big)
\end{equation}
and\\[10pt]
$\displaystyle   \{\Pi_{IJ}(x), \Pi^i{}_{KL}(y)\}=-\frac\alpha2\, \delta(x-y)\,\Big(\epsilon_{KLIP4}\cP^i{}^P{}_J(x) $\\[5pt]
$\displaystyle- \epsilon_{KLJP4}\cP^i{}^P{}_I(x)\Big)+\frac\alpha4\, \delta(x-y)\,\Big(\eta_{IL}\epsilon_{JKMN4}$\\[5pt]
$\displaystyle-\eta_{JL}\epsilon_{IKMN4}-\eta_{IK}\epsilon_{JLMN4} +\eta_{JK}\epsilon_{ILMN4}\Big)\cP^{iMN}(x)$\\[5pt]
$\displaystyle
+\frac12\, \delta(x-y)\,\Big(\eta_{IL}\Pi^i{}_{JK}(x) -\eta_{JL}\Pi^i{}_{IK}(x)$
\begin{equation}\label{15}-\eta_{IK}\Pi^i{}_{JL}(x) +\eta_{JK}\Pi^i{}_{IL}(x)\Big)\, .
\end{equation}
It is worth noticing  that in the topological limit $\alpha=0$ all
the constraints are first class. This observation leads to the
following, apparent puzzle. Namely, as we said above the kinematical
phase space is 60 dimensional. On the other hand for $\alpha=0$ we
have $10 + 30$ first class constraints that remove from this phase
space 80 degrees of freedom. How is this possible? To answer this
let us notice that not all the constraints are independent. Indeed
taking the covariant divergence of the $\Pi^i{}_{IJ}$ constraint and
making use of the Bianchi identity we see that
\begin{equation}\label{16}
    (\cD_i \Pi^i)_{IJ}=-\beta \Pi_{IJ}
\end{equation}
and thus the set of constraints is reducible. It follows that we
have only 30 independent first class constraints $\Pi^i{}_{IJ}$,
which remove exactly 60 dimensions from the phase space, as it
should be since the theory with $\alpha=0$ is topological.

Returning to the case $\alpha\neq0$ we notice that the action
(\ref{1}) is invariant under local gauge transformations that belong
to the Lorentz subgroup $\SO(3,1)$ of the initial de Sitter group
$\SO(4,1)$\footnote{In what follows we restrict ourself to the
positive cosmological constant case; the negative cosmological
constant and the Anti de Sitter group $\SO(3,2)$ can be analyzed
analogously.}. It follows that it is natural to expect that one can
simplify the algebra of constraints (\ref{13})--(\ref{15}) if one
decomposes the constraints into that belonging to the Lorentz and
the translational parts of the algebra. From (\ref{10}) we get
\begin{equation}\label{12.1}
    \Pi_{\alpha 4}(x) \equiv \Pi_{\alpha}(x) = \left(\cD^\omega_i \cP^i\right)_{\alpha 4}(x) - \frac{1}{\ell}e_i{}^\beta{}(x)\cP^i{}_{\alpha\beta}(x)\approx0
\end{equation}
$
\Pi_{\alpha\beta}(x) = \left(\cD^\omega_i \cP^i\right)_{\alpha\beta}(x) $\begin{equation}\label{12.2} -\frac{1}{\ell} e_i{}_\alpha{}(x)\cP^i{}_{\beta 4}(x)+\frac{1}{\ell}e_i{}_\beta{}(x)\cP^i{}_{\alpha 4}(x)\approx0
\end{equation}
while from (\ref{11})
\begin{equation}\label{12.3}
\Pi^i{}_{\alpha 4}(x) \equiv \Phi^i{}_{\alpha}(x)= \left(2\epsilon^{ijk}\, F_{jk}{}_{\alpha 4}(x)-\beta\, \cP^{i}_{\alpha 4}(x) \right)\approx0\quad\quad
\end{equation}\\
$
\Pi^i{}_{\alpha\beta}(x)\equiv \Phi^i{}_{\alpha\beta}(x)=
$\begin{equation}\label{12.4} = \Big(2\epsilon^{ijk}\, F_{jk}{}_{\alpha\beta}(x)-\beta\, \cP^{i}_{\alpha\beta}(x)- \frac\alpha2\, \epsilon_{\alpha\beta\gamma\delta}\, \cP^i{}^{\gamma\delta}(x)\Big)\approx0
\end{equation}

One then finds that  the algebra of constraints (\ref{14}),
(\ref{15})  simplifies a lot, and the only brackets that do not
vanish weakly are
\begin{equation}\label{17}
    \left\{\Pi_\alpha(x), \Phi^i{}_{\gamma\delta}(y)\right\}\approx-\frac\alpha2\, \delta(x-y)\, \epsilon_{\gamma\delta\alpha\rho}\, \cP^i{}^\rho{}_4(x)
\end{equation}
\begin{equation}\label{18}
    \left\{\Pi_\alpha(x), \Phi^i{}_{\gamma}(y)\right\}\approx-\frac\alpha4\, \delta(x-y)\, \epsilon_{\alpha\gamma\rho\sigma}\, \cP^i{}^{\rho\sigma}(x)
\end{equation}
$$
   \{\Phi^i{}_{\alpha}(x), \Phi^j{}_{\beta\gamma}(y)\} \approx
2\alpha\, \epsilon^{ijk} \delta(x-y)\, \epsilon_{\alpha\beta\gamma\delta}\, A_k{}^\delta{}_4(x) =$$ \begin{equation}\label{19}\frac{2\alpha}\ell\, \epsilon^{ijk} \delta(x-y)\, \epsilon_{\alpha\beta\gamma\delta}\, e_k{}^\delta(x)
\end{equation}

Now we can turn to the next step of canonical analysis, i.e.,  to
checking if there are any tertiary constraints. The Hamiltonian,
being a combination of constraints reads
\begin{equation}\label{20}
    H= -2 A^\alpha\, \Pi_\alpha-A^{\alpha\beta}\, \Pi_{\alpha\beta}-2B_i{}^{\alpha}\, \Phi^i{}_{\alpha}-B_i{}^{\alpha\beta}\, \Phi^i{}_{\alpha\beta}
\end{equation}
It follows from (\ref{17}-\ref{19}) that we have to satisfy the
following conditions to ensure that the constraints are preserved by
time evolution, generated by hamiltonian (\ref{20})
\begin{equation}\label{21}
    \dot \Pi_\alpha = \frac{\alpha}{2} \big( B_i{}^{\beta}\, \cP^i{}^{\gamma\delta} + B_i{}^{\beta\gamma}\, \,\cP^i{}^{\delta}{}_4\big)\epsilon_{\alpha\beta\gamma\delta} \approx0
\end{equation}
\begin{equation}\label{22}
\dot \Phi^i{}_\alpha  =- \alpha\,\big( \frac{2}{\ell}\epsilon^{ijk}\,   B_j{}^{\beta\gamma}\, e_k{}^{\delta}{} -\frac{1}{2} A^\beta \mathcal{P}^{i\,\gamma\delta}\big)\epsilon_{\alpha\beta\gamma\delta}\approx0
\end{equation}
\begin{equation}\label{23}
\dot \Phi^i{}_{\alpha\beta}   = -\alpha\,  \big(\frac{4}{\ell}\epsilon^{ijk}\, B_j{}^{\gamma}\, e_k{}^{\delta}{}+A^\gamma \mathcal{P}^{i\,\delta}_{~~~4}\big)\epsilon_{\alpha\beta\gamma\delta}\approx0
\end{equation}

These equations can be solved for Lagrange multipliers (we have 34 equations for 34 unknowns $B_i{}^{\beta}$, $B_i{}^{\beta\gamma}$, $A^\gamma$ with arbitrary coefficients) and thus
there are no tertiary constraints.

Notice however that there is an ambiguity in Dirac procedure in the case of diff-invariant systems, i.e., such that hamiltonian is a combination of constraints. The usual approach is to check if one can solve the vanishing of time derivative of the constraints condition for Lagrange multipliers, as we did above. But this is, clearly, not a general solution of these conditions. In general one may look for the solutions with arbitrary values of the Lagrange multipliers, but instead restricting the phase space (for example if we impose the condition that all the Lagrange multipliers in (\ref{21})--(\ref{23}) are arbitrary there would be additional constraints saying that components of tetrad and momenta are to be equal zero.) Notice that this problem does not arise in the case of the hamiltonian not being weakly zero, because then the resulting  equations pertaining to the time invariance of the constraints are non-homogeneous. Thus the procedure that is usually employed does not seem to provide a complete characterization of the phase space, but we will adopt it here, leaving the discussion of this subtle point to the future work.

\section{Simplifying the constraints}

The aim of this section is to rewrite the system of constraints
(\ref{13})--(\ref{15}) in a form that makes comparison with
constraints of General Relativity with Holst term, discussed in
\cite{Holst:1995pc}. In what follows we will borrow some ideas from
the paper of Perez and Rezende \cite{Rezende:2009sv}. (Similar ideas, albeit in more restricted setting, were discussed, e.g., in \cite{Randono:2008bb} and \cite{Date:2008rb}.)

In the first step let us rearrange the constraints
(\ref{13})--(\ref{15}) to write them in the following form
\begin{eqnarray}\label{r1a}
\Phi^i_\alpha{~}&=&\cP^{i}_{\alpha} - \frac{4}{\ell\beta}\epsilon^{ijk}\, \cD^\omega_j e_k{}_{\,\,\alpha}\approx 0\\
\Phi^i_{\alpha\beta}&=&\cP^{i}_{\alpha\beta}-M_{\alpha\beta}{}^{\gamma\delta}\, F^{}_{jk}{}_{\,\gamma\delta}\,\epsilon^{ijk} \approx 0\\
\Pi_{\alpha\beta}&=& \frac{2}{\ell^2}\,\epsilon^{ijk} \cD^\omega_i \Big(K_{\alpha\beta}{}^{\gamma\delta}\,e_{j\,\gamma}e_{k\,\delta}\Big)\approx 0\\
\Pi_{\alpha}{~}&=& \frac{1}{\ell}\,\epsilon^{ijk}\,K_{\alpha\beta}{}^{\gamma\delta}\, e^{\,\,\beta}_{i}\,R_{jk\,\,\gamma\delta}\nonumber\\
& &-\frac{2\alpha}{(\alpha^2+\beta^2)\ell^3}\,\epsilon^{ijk}\,
\epsilon_{\alpha\beta\gamma\delta}\;e^{\;\beta}_{i}\,e^{\;\gamma}_{j}\,e^{\;\delta}_{k}\approx 0\label{r1z}
\end{eqnarray}
Recall that the coupling constant $\alpha$ and $\beta$ satisfy the
identity $\alpha/(\alpha^2+\beta^2)=\ell^2 /G$, while the operators
$M$ and $K$ are defined to be
\begin{eqnarray}
M^{\alpha\beta}{}_{\gamma\delta}&\equiv& \frac{\alpha}{(\alpha^2+\beta^2)}( \gamma\, \delta^{\alpha\beta}_{\gamma\delta}-\epsilon^{\alpha\beta}{}_{\gamma\delta}),\\
K^{\alpha\beta}{}_{\gamma\delta}&\equiv&
\frac{\alpha}{(\alpha^2+\beta^2)}(\frac{1}{\gamma}\,
\delta^{\alpha\beta}_{\gamma\delta}+\epsilon^{\alpha\beta}{}_{\gamma\delta})\,
.
\end{eqnarray}
Also recall that the action (\ref{1}), after solving for $B$ and
expressing the resulting action in terms of the
$\SO(3,1)$-connection $\omega$ and tetrad $e$, has the form
\cite{Freidel:2005ak}
\begin{eqnarray}
S&=&\frac{1}{G}~\int\epsilon^{\alpha\beta\gamma\delta}(R_{\mu\nu\,\alpha\beta}\,e_{\rho\,\gamma}e_{\sigma\,\delta}
-\frac{\Lambda}{3}e_{\mu\,\alpha}e_{\nu\,\beta}e_{\rho\, \gamma}e_{\sigma\, \delta})\epsilon^{\mu\nu\rho\sigma}\nonumber\\
&+&\frac{2}{G\gamma}\int R_{\mu\nu\,\alpha\beta}\,e_{\nu}^{\,\alpha}e_{\rho}^{\, \beta}\,\epsilon^{\mu\nu\rho\sigma}\nonumber\\
&+&\frac{\gamma^2+1}{\gamma\, G} NY_4
+\frac{3\gamma}{2G\Lambda}P_4-\frac{3}{4G\Lambda}
E_4\label{Y}\, .
\end{eqnarray}
One immediately recognizes here the standard gravitational action in
the first line, and the Holst term, whose strength is governed by the
Immirzi parameter \mbox{$\gamma=\beta/\alpha$} in the second.
The last three terms are proportional to topological invariants
(Nieh-Yan, Ponryagin, and Euler):
\begin{eqnarray}
& & NY_4=\int (T_{\mu\nu\,\alpha}T^{\alpha}_{\rho\sigma}-2\,R_{\mu\nu\,\alpha\beta}e_{\nu}^{\,\alpha}e_{\rho}^{\, \beta})\,\epsilon^{\mu\nu\rho\sigma}\, ,\label{ny}\\
& &  P_4=\int R_{\mu\nu\,\alpha\beta}R^{\alpha\beta}_{\rho\sigma}\,\epsilon^{\mu\nu\rho\sigma} ,\label{po}\\
& &E_4=\int R_{\mu\nu\,\alpha\beta}R_{\rho\sigma\,\gamma\delta}\epsilon^{\alpha\beta\gamma\delta}\,\epsilon^{\mu\nu\rho\sigma}\, .\label{eu}
\end{eqnarray}

As we will show, in the case when the constant time surface is
without boundaries $\partial \Sigma=0$, the topological terms play
the role of the generating functional for canonical transformations,
which simplify the constraints considerably \cite{Rezende:2009sv}.
The key observation is that Pontryagin and Nieh-Yan invariants can
be expressed as total derivatives
\begin{eqnarray}
&&    NY_4=4\int \partial_\mu \Big( e_{\nu\;\alpha} \cD^\omega_\rho e^{\;\alpha}_{\sigma} \Big)\,\epsilon^{\mu\nu\rho\sigma}\\
 &&P_4= 4\int \partial_\mu \Big(\omega_{\nu\;ab}\,\partial_\rho \omega_\sigma^{ab}+\frac{2}{3}\omega_{\nu\;ab} \,\omega_{\rho\;\,c}^{\;a} \,\omega_{\sigma}^{cb}\Big)\,\epsilon^{\mu\nu\rho\sigma}\qquad\end{eqnarray}
The same holds for Euler class. However in this case one has to make
use of self and anti-self dual combinations of Lorentz connection
\begin{equation}
{^{\pm}} \omega_{i}^{\alpha\beta} = {\frac{1}{2}}(\omega_{i}^{\alpha\beta} \mp {\frac{i}{2}}
\epsilon^{\alpha\beta}_{ \ \ \gamma\delta} \omega_{i}^{\gamma\delta}),\quad      {^{\pm}
} \omega_i^{\gamma\delta}\,\epsilon^{\alpha\beta}_{ \ \ \gamma\delta}  = \pm i {^{\pm}} \omega^{\alpha\beta}_i
\end{equation}
and curvature (see e.g., \cite{GarciaCompean:1999kj})
\begin{equation}
{}^{\pm}R^{\alpha\beta}_{\mu\nu}=\frac{1}{2}\Big( R_{\mu\nu}^{\alpha\beta}\mp\frac{i}{2}\epsilon^{\alpha\beta}_{\;\;\gamma\delta }R_{\mu\nu}^{\gamma\delta} \Big)\, .
\end{equation}
It can be checked that both Pontryagin and Euler class can be
rewritten with the help of ${}^{\pm} R^{\alpha\beta}_{\mu\nu}$ as follows
\begin{equation}
 P_4 = \int \epsilon^{\mu\nu\sigma\rho}\, ( {}^{+} R^{\alpha\beta}_{\mu\nu} {}^{+} R_{\rho\sigma\;\alpha\beta}+ {}^{-} R^{\alpha\beta}_{\mu\nu} {}^{-} R_{\rho\sigma\;\alpha\beta})
\end{equation}
\begin{equation}
 E_ 4 = 2i \int \epsilon^{\mu\nu\sigma\rho}\, ( {}^{+} R^{\alpha\beta}_{\mu\nu} {}^{+} R_{\rho\sigma\;\alpha\beta}- {}^{-} R^{\alpha\beta}_{\mu\nu} {}^{-} R_{\rho\sigma\;\alpha\beta})
\end{equation}
Introducing
 \begin{equation}
\cC ^\mu (\omega)=\Big(\omega_{\nu\;\alpha\beta}\,\partial_\rho \omega_\sigma^{ab}+\frac{2}{3}\omega_{\nu\;ab} \,\omega_{\rho\;\,c}^{\;a} \,\omega_{\sigma}^{cb}\Big)\,\epsilon^{\mu\nu\rho\sigma}
\end{equation}
we write Pontryagin and Euler classes as total derivatives
\begin{eqnarray}
  P_4&=&4\int \Big(\partial_\mu \cC ^\mu (^+\omega)+ \partial_\mu \cC ^\mu (^-\omega) \Big) \\
E_4 &=&8i \int\Big(\partial_\mu \cC ^\mu (^+\omega)-\partial_\mu \cC ^\mu (^-\omega)\Big)
\end{eqnarray}
Therefore the topological part of action (\ref{Y}) takes the form
\begin{eqnarray}
    S_T&=& \frac{4}{\beta\ell^2}  \int \partial_\mu \big( e_{\nu\;\alpha} \cD^\omega_\rho e^{\;\alpha}_{\sigma} \big)\,\epsilon^{\mu\nu\rho\sigma} \nonumber \\
&+&\frac{2\alpha}{(\alpha^2+\beta^2)}\frac{\beta}{\alpha}\int\partial_\mu\Big( \cC ^\mu (^+\omega)+ \cC ^\mu (^-\omega) \Big) \nonumber\\
&-&i \frac{2\alpha}{(\alpha^2+\beta^2)}\int\partial_\mu\Big( \cC ^\mu (^+\omega)- \cC ^\mu (^-\omega)\Big)\, .
\end{eqnarray}
It is worth noticing that in spite of the presence of the imaginary
$i$ here, the action $S_T$ is real (for real $\gamma$.)

For constant time surfaces, being a manifold without boundary
($\partial \Sigma=0$), all total spacial derivatives terms drop out
and only the ones with total time derivative survive
$$
  S_T=\int \partial_0  W(e,\omega)\, ,
$$
where $W(\omega,e)$ is a functional of torsion and self and
anti-self dual Chern-Simons forms $\cL_{CS}\equiv \cC^0$
\begin{eqnarray}
    & &  W(e,\omega)=\frac{4}{\beta\ell^2} \int_\Sigma  \,\epsilon^{ijk} \, \big( e_{i\;\alpha} \cD^\omega_j e^{\;\alpha}_{k} \big)+\\
&+& \frac{2\alpha}{(\alpha^2+\beta^2)} \int_\Sigma \Big((\gamma-i)\cL_{CS} (^+\omega)+(\gamma+i) \cL_{CS} (^-\omega) \Big).\nonumber\label{Wfunctional}
\end{eqnarray}
Having the functional $W$ we can make canonical transformation,
which defines new momenta $\mathscr{P}^i_a$, $\mathscr{P}^i_{ab}$ of the
tetrad $e$ and the connection $\omega$, respectively
\begin{equation}\label{nm}
  \mathscr{P}^i_\alpha=  \cP^i_\alpha+\{\cP^i_\alpha, W(\omega,e)\}, \quad   \mathscr{P}^i_{\alpha\beta}=  \cP^i_{\alpha\beta}+\{\cP^i_{\alpha\beta}, W(\omega,e)\}
\end{equation}
with
\begin{equation}\label{np}
 \left\{e_i^\alpha,\mathscr{P}^j_\beta\right\} =\frac{1}{2}\ell\, \delta^j_i\, \delta^\alpha_\beta\quad \mathrm{and}\quad       \left\{\omega_i^{\alpha\beta},\mathscr{P}^j_{\gamma\delta}\right\} =\frac{1}{2}\delta^j_i\,\delta^{\gamma\delta}_{\alpha\beta}\,.
\end{equation}
Since the variations of the functional $W(\omega,e)$ are
\begin{eqnarray}
       \frac{1}{2} \frac{\delta W}{\delta \omega^{\alpha\beta}_{i}}&=&M_{\alpha\beta}{}^{\gamma\delta}\, R^{}_{jk}{}_{\,\gamma\delta}\,\epsilon^{ijk}-\frac{4}{\beta\ell^2}\,e_{j\;\alpha}\, e_{k\;\beta}\,\epsilon^{ijk}~~~~~~\\
    \frac{1}{2}\frac{\delta W}{\delta e_i^\alpha}&=& \frac{4}{\ell\beta}\epsilon^{ijk}\, \cD^\omega_j e_k{}_{\,\,\alpha}
\end{eqnarray}
we find that the resulting constraints, expressed in terms of new
momenta (\ref{nm}) take the form
\begin{eqnarray}
\Phi^i_\alpha{~}&=&  \mathscr{P}^i_\alpha\approx 0,\label{r1x}\\
\Phi^i_{\alpha\beta}&=&\mathscr{P}^i_{\alpha\beta}- \frac{2}{\ell^2}K^{~~~\gamma\delta}_{\alpha\beta} \,e_{j\;\gamma}\,e_{k\;\delta}\,\epsilon^{ijk}\approx 0\label{r1y}\\
\Pi_{\alpha\beta}&=& \frac{2}{\ell^2}\,\epsilon^{ijk}K_{\alpha\beta}{}^{\gamma\delta}\, \cD^\omega_i \Big(e_{j\,\gamma}e_{k\,\delta}\Big)\approx 0\label{r1c}\\
\Pi_{\alpha}{~}&=& \frac{1}{\ell}\,\epsilon^{ijk}\,K_{\alpha\beta}{}^{\gamma\delta}\, e^{\,\,\beta}_{i}\,F_{jk\,\,\gamma\delta}\approx 0\label{r1d}
\end{eqnarray}
This form of constraints will be our starting point in checking
equivalence with the ones proposed by Holst \cite{Holst:1995pc},
which will prove, in turn, that they describe General Relativity, as
expected. To establish this equivalence we will have to fix the time
gauge. We will turn to this problem  in the next section.

Before closing this section let us make an important remark. The
considerable simplification of the constraints relies heavily on the fact
that the constant time surfaces are manifolds without boundary. In
the case when  boundaries are present the analysis of the
constraints becomes much more involved. We will address this issue
in the forthcoming paper.

\section{Time gauge}
In order to make contact with the Hamiltonian analysis of Holst, we have
to fix the gauge so as to remove the time component of the tetrad
and then to relate momenta associated with Lorentz connection
with an appropriate combination of the remaining tetrad components.

To this end, let us introduce the gauge condition which must be added to the list of constraints
\begin{equation}\label{gauge}
    e^0_i \approx 0
\end{equation}
In this gauge  the constraints $\mathscr{P}^i_0\approx 0$ can be
removed by turning to the Dirac bracket so the remaining
constraints (\ref{r1x}), (\ref{r1y}) take the form
\begin{eqnarray}
& &\mathscr{P}^{i}_{a}\approx 0\label{v1}\\[7pt]
& &\mathscr{P}_{0a}^i+\frac{2\alpha}{\ell^2(\alpha^2+\beta^2)}\epsilon^{ijk}\epsilon_{abc}e^b_j\,e^c_k\approx 0\label{v2}\\
& &\mathscr{P}_{ab}^i-\frac{2\alpha}{\ell^2(\alpha^2+\beta^2)}\frac{1}{\gamma}\epsilon^{ijk}\delta^{cd}_{ab}\,e_{j\;c}\,e_{k\;d}\approx 0\label{v3}
\end{eqnarray}
where we have used the convention $\epsilon^{0abc}=\epsilon^{abc}$ and $\epsilon_{0abc}=-\epsilon_{abc}$.

Combining the last two equations we find constraints for generalized
self and anti self-dual parts of $\mathscr{P}$
\begin{eqnarray}
  & &{}^+\mathds{P}_{a}^i \approx 0\label{v4}\\
& & {}^-\mathds{P}_{a}^i
+\frac{4\alpha}{(\alpha^2+\beta^2)\ell^2}\epsilon^{ijk}\epsilon_{abc}e^b_j\,e^c_k\approx
0\label{v5}
\end{eqnarray}
where we define
\begin{equation}\label{v6}
  {}^\pm\mathds{P}_{a}^i=   \mathscr{P}_{0a}^i \pm \frac{\gamma}{2}\,\epsilon_{abc} \mathscr{P}^{i\;bc}
\end{equation}
It can be easily checked that ${}^\pm\mathds{P}$ are momenta
associated with generalized (anti) self-dual combinations of Lorentz
connection (which for $\gamma =\pm i$ become usual self and
anti self dual ones)
 \begin{equation}\label{v7}
  {}^\pm w^{a}_i =\omega^{0a}_i  \pm\,  \frac{1}{2\gamma}\,\epsilon^{abc} \omega_{i\;bc}
\end{equation}
with the Poisson brackets being
\begin{equation}
     \{ {}^\mp w^{a}_i ,  {}^\pm\mathds{P}_{d}^j \}=0\, ,\qquad \{ {}^\pm w^{a}_i ,  {}^\pm\mathds{P}_{d}^j \}=\delta^j_i \delta^a_d\, .
\end{equation}

Let us now turn to the constraint $\Pi_{\alpha\beta}$ (\ref{r1c}). Decomposing it into components we find
\begin{eqnarray}
& &\Pi_{ab}= \frac{2\alpha}{(\alpha^2+\beta^2)\ell^2}\,\epsilon^{ijk}\Big(-2\epsilon_{abc}\,\omega_{i}^{0\;d}\, e_{j\,d}\, e_{k}^{\;c}+\nonumber\\
&&+\frac{2}{\gamma}\,\big(\partial_i(e_{j\,a}\,e_{k\,b})
+\omega_{i\,a}^{~~c}\,e_{j\,c}\,e_{k\,b}-\omega_{i\,b}^{~~c}\,e_{j\,c}e_{k\,a}) \Big)\label{v8}\\
& &\Pi_{0a}=\frac{-2\alpha}{(\alpha^2+\beta^2)\ell^2}\,\epsilon^{ijk}\,\nonumber\\
&&\Big(\epsilon_{abc}\,\big(\partial_i(e_j^b\,e^c_k)+2\omega_{i}^{bd}\,e_{j\,d}\,e^c_k) +\frac{2}{\gamma}\,\omega_{i~0}^{~~~b}\, e_{j\,b}\, e_{k\,a}\Big)\qquad\label{v9}
\end{eqnarray}
Taking the combination $\displaystyle\Pi_{0a}\pm\frac{\gamma}{2}\epsilon_{abc}\Pi^{bc}$ we get
\begin{equation}    \frac{4\alpha}{(\alpha^2+\beta^2)\ell^2}\,\epsilon^{ijk}\,\Big(\frac{1+\gamma^2}{\gamma}\Big)\,\omega_{i\,0}^{~~b}\, e_{j\,a}\, e_{k\,b} \approx 0 \label{v10}
\end{equation}
and
\begin{eqnarray}
&&\frac{4\alpha}{(\alpha^2+\beta^2)\ell^2}\,\epsilon^{ijk}\,
\Big(\big(\frac{1-\gamma^2}{\gamma}\big)\,\omega_{i\,0}^{~~b}\, e_{j\,a}\, e_{k\,b}\qquad\qquad\qquad\qquad\nonumber\\
&&\qquad\qquad\qquad-\epsilon_{abc}\,\big(\partial_i(e_j^b\,e^c_k)+2\omega_{i}^{bd}\,e_{j\,d}\,e^c_k)\Big)\approx 0\label{v11}\, .
\end{eqnarray}
From these two equations it follows that\footnote{This is obvious for $\gamma^2\neq-1$. For $\gamma^2=-1$ eq. (\ref{v10}) is identically satisfied, but then, since the connection is real, the real and imaginary parts of (\ref{v11}) lead to (\ref{v12}), (\ref{v13}).}
\begin{eqnarray}
&&\qquad\qquad   \omega_{i~0b}^{}\, e_{j\,a}\,e^b_k \, \epsilon^{ijk}\approx 0\label{v12}\\
&&\Big(\partial_i(e_j^b\,e^c_k)
+2\omega_{i~d}^{b}\,e_j^d\,e^c_k\Big)\epsilon^{ijk}\,\epsilon_{abc}\approx 0\label{v13}
\end{eqnarray}
which expressed in terms of the new variables
\begin{equation}
 \omega^{0a}_i= \frac{1}{2}  \big( {}^+ w^{a}_i+{}^- w^{a}_i\big)\, ,\quad  \omega^{ab}_i= \frac{\gamma}{2} \epsilon^{abc} \big( {}^+ w_{i\,c}-{}^- w_{i\,c}\big)\label{v14}
\end{equation}
take the form of the Gauss and the boost constraints
\begin{equation}
G_a\equiv({}^+ w^{b}_i+ {}^- w^{b}_i)\, e_{j\,a}\, e_{k\,b} \, \epsilon^{ijk}\approx 0\, ,\label{v15}
\end{equation}
\begin{equation}
B_a\equiv\Big(\partial_i(e_j^b\,e^c_k)\,\epsilon_{abc}-\gamma\,({}^+ w^{b}_i- {}^- w^{b}_i)\,e_{j\,a}\,e_{k\,b}\Big)\epsilon^{ijk}\approx 0\, .\label{v16}
 \end{equation}

We can handle the scalar part of (\ref{r1d})
\begin{equation}
\mathsf{S}= \frac{\alpha}{(\alpha^2+\beta^2)\ell}\,\Big( \frac{2}{\gamma}\,e_{i}^{c}\,R_{jk\,0c}-\epsilon_{abc}\, e^{\,a}_{i}\,F_{jk}^{\,bc}\big)\,\epsilon^{ijk}\approx 0\label{v17}
\end{equation}
similarly, obtaining as a result the expression
$$
\Big[ \Big(\frac{1+\gamma^2}{\gamma}\Big)\epsilon^{dbc} \partial_j ({}^+ w_{k\,d})+\Big(\frac{1-\gamma^2}{\gamma}\Big)\epsilon^{dbc}\partial_j({}^- w_{k\,d})$$
$$+(1+\gamma^2){}^+ w_{j}^b {}^- w_{k}^c -\Big(\frac{1+\gamma^2}{2}\Big) {}^+ w_{j}^b {}^+ w_{k}^c-\Big(\frac{3-\gamma^2}{2}\Big) {}^- w_{j}^b {}^- w_{k}^c$$
$$
-\frac{2}{\ell^2}e^{\,b}_{j}\,e^{\,c}_{k} \Big]\epsilon_{abc}\, e^{a}_{i}\,\epsilon^{ijk}\, \frac{\alpha}{(\alpha^2+\beta^2)\ell}\approx 0\, .$$
As for the vector part of (\ref{r1d})
\begin{equation}
\mathsf{V}_a=\frac{\alpha}{(\alpha^2+\beta^2)\ell}\,\Big( \frac{2}{\gamma}\,e_i^{\,b}\,R_{jk\,ab}+2\epsilon_{ab~}^{~~c}\, e^{b}_{i}\,R_{jk\,0c}\Big)\,\epsilon^{ijk}\approx 0\, ,\label{v18}
\end{equation}
we find
$$
\Big[2\epsilon_{abc} \partial_j^- w_{k}^a-\Big(\frac{1+\gamma^2}{2\gamma}\Big) {}^+ w_{j\,b} {}^+ w_{k\,c}-\Big(\frac{3\gamma^2-1}{2\gamma}\Big) {}^- w_{j\,b} {}^- w_{k\,c}$$
$$-\Big(\frac{1+\gamma^2}{2\gamma}\Big)({}^+ w_{j\,b} {}^- w_{k\,c}+{}^- w_{j\,b} {}^+ w_{k\,c})
\Big]\,e^{c}_{i}\,\epsilon^{ijk}\, \frac{2\ell}{G}\approx 0\, .$$
It should be noted that in the case $\gamma^2=-1$  all terms containing ${}^+ w_{i\,a}$ cancel.
Notice also that with the help of the Gauss constraint the vector constraint can be reduced to the form
$$
\Big[2\epsilon_{abc} \partial_j^- w_{k}^a+\Big(\frac{1-\gamma^2}{\gamma}\Big) {}^- w_{j\,b} {}^- w_{k\,c}\Big]\,e^{c}_{i}\,\epsilon^{ijk}\approx 0\, .$$

The form of the constraints that we have obtained so far is still not the final one. At some point we will have to get rid of the constraint ${}^+\mathds{P}_{a}^i \approx 0$ and eliminate the dependence on ${}^+ w_{j\,b}$ of all the remaining constraints. However in order to be able to do that we must simplify the form of the boost constraint (\ref{v16}). To see how this can be done we multiply the constraint  (\ref{v4}) by tetrad and decompose the resulting constraint $C_{ab}$ into symmetric and antisymmetric parts
\begin{equation}
C_{(ab)}= {}^+\mathds{P}_{i\,(a}  e_{b)}^i\;,\qquad C_{[ab]} ={}^+\mathds{P}_{i\,[a}  e_{b]}^i\, .\label{v19}
\end{equation}
Let us now calculate the Poisson bracket of $C_{ab}$ with the scalar constraint
$$
   \{ C_{ab}, S\} =
   $$
   \begin{equation}
   \frac{(1+\gamma^2)\ell}{\gamma G}\epsilon^{ijk}e_{j\,b}\Big(2\partial_k e_{i\,a }-\gamma ({}^+ w^{c}_k-{}^- w^{c}_k)e_i^d\epsilon_{acd}\Big)\, .\label{v20}
\end{equation}
The bracket of the antisymmetric part  $C_{[ab]}$ gives exactly the boost constraint\footnote{Instead of $C_{[ab]} ={}^+\mathds{P}_{i\,[a}  e_{b]}^i$ just take the expression $C_{ab}\,\epsilon^{abc} ={}^+\mathds{P}_{i\,a}\,e_{b}^i\,\epsilon^{abc}$, so
$\{ C_{ab}\,\epsilon^{abc} , S\}=\frac{(1+\gamma^2)\ell}{\gamma G}B^c$.} (\ref{v16}). However, the bracket of the symmetric part $C_{(ab)}$  leads to the secondary constraint
\begin{eqnarray}
&&B_{ab}\equiv 2\epsilon^{ijk}(e_{j\,b}\partial_k e_{i\,a }+e_{j\,a}\partial_k e_{i\,b })\nonumber\\
&&-\gamma ({}^+ w^{c}_k-{}^- w^{c}_k)\,e_i^d\,(e_{j\,b}\epsilon_{acd}+e_{j\,a}\epsilon_{bcd})\approx 0\, .\label{v21}
\end{eqnarray}
Clearly, this constraint would arise if we impose the requirement that all the constraints are to be preserved in the time evolution. Therefore one has to add $B_{ab}$ to the set of constraints of the theory. But then the road suddenly becomes sunny. It suffices to note that the boost constraints $B_a$ and the newly derived constraints $B_{ab}$ are just the antisymmetric and symmetric parts of the simple constraint
    \begin{equation}\label{v25}
\Big(2\Gamma^c_i-\gamma\,({}^+ w_{i}^c- {}^- w_{i}^c)\Big)\approx 0\, ,
\end{equation}
where $\Gamma^a_i$ is a (unique) solution of the Cartan first structural equation
\begin{equation}\label{v23}
 \Big(\partial_{[i} e_{j]\,b})
+\epsilon_{bcd}\Gamma_{[i}^{c}\,e_{j]}^{d}\Big)=0\, .
\end{equation}

Using (\ref{v25}) and (\ref{v5}) we get rid of both ${}^+\mathds{P}_{a}^i $ and  ${}^+ w_{j\,b}$, replacing ${}^+ w_{j\,b}$ in all the remaining constraints with the solution of (\ref{v25}). Similarly using (\ref{v1}) and (\ref{v5}) we can identify the momentum of ${}^- w_{j\,b}$ with
$$
-\frac{4\alpha}{(\alpha^2+\beta^2)\ell^2}\,\epsilon^{ijk}\,\epsilon_{abc}\,e^b_j\,e^c_k\, .
$$
What remains are therefore $3$ Gauss, $3$ vector and $1$ scalar constraints, all of them first class, constraining the $18$-dimensional phase space of ${}^- w_{j\,b}$ and its momenta. Thus the dimension of physical phase space is $18-14=4$ as it should. Of course, the final set of constraints we have obtained has exactly the form of the constraints describing gravity, cf.\ \cite{Holst:1995pc}. This completes our analysis of the canonical structure of $\SO(4,1)$ constrained BF theory.

\section{Comments on quantization}

Let us conclude this paper with some comments concerning quantization. Clearly, one can take the first class Gauss, vector, and scalar constraints as a starting point in construction of the quantum theory, as it is done in Loop Quantum Gravity \cite{Rovelli:2004tv}, \cite{Thiemann:2007zz}. However, the structure of constraints of the original theory opens another possibility of devising a perturbative expansion in parameter $\alpha$ around topological vacuum. Here we will describe briefly this perturbative theory leaving details to a separate publication.

Our starting point will be the set of constraints (\ref{r1a})--(\ref{r1z}). Consider now the canonical transformation (\ref{nm}). Its quantum counterpart can be easily found. To see how, take the mechanical model in which one makes the transformation (see \cite{Mercuri:2007ki})
$$
p_i \rightarrow p'_i = p_i +\left\{p_i, f(q)\right\}
$$
so that quantum mechanically we have
$$
\hat p_i \rightarrow \hat p'_i = \hat p_i +i\left[\hat p_i, f(\hat q)\right] \, .
$$
If we represent $\hat p_i=i\partial/\partial q^i$ then $\hat p'_i = i\partial/\partial q^i -\partial f(q)/\partial q^i$. Therefore if we decompose the wave function $\psi(q) = \exp(-i f(q))\, \psi'(q)$ then
$$
\hat p' \psi(q) =\exp(-i f(q))\,\hat p'  \psi'(q)\, ,
$$
which means that we just have to multiply the wave function with the phase $\exp(-i f(q))$ and then use the standard representation of the new momenta $p'$ as the derivatives over positions. In the case at hands (\ref{nm}), it is therefore sufficient to multiply the wave function by the prefactor $\exp\left(-iW(e,\omega)\right)$ where $W(e,\omega)$ is given by (\ref{Wfunctional}), and replace all the momenta $\cP$ with the new ones $\mathscr{P}$. Then we can just use the constraints (\ref{r1x})--(\ref{r1d}).

When $\alpha=0$ these constraints reduce to the first class set
\begin{equation}\label{6.1}
  \mathscr{P}^i_\alpha\approx 0\, , \quad \mathscr{P}^i_{\alpha\beta}\approx 0\, .
\end{equation}
The wave function annihilated by them is just a constant, and thus the full physical wave function is a phase $\exp\left(-iW(e,\omega)\right)$. Clearly, and not surprisingly, in this case the wave function is the Kodama state \cite{Kodama:1990sc} (strictly speaking this is the Kodama state for $\SO(4,1)$ multiplied by the phase proportional to Euler class of a constant time manifold.) Notice that here this state is delta function normalizable, because all our constraints are real (cf.\ \cite{Freidel:2003pu}). The simplicity of the zeroth order (in $\alpha$) solution reflects the fact that to this order the theory is topological.

Let us now turn to devising the $\alpha$ perturbative theory. The constraints (\ref{r1x})--(\ref{r1d}) all have the form $\Phi = \Phi^{(0)} + \alpha\, \Phi^{(1)}$ (for the last two $\Phi^{(0)}=0$). We also expand the wave function in the series in $\alpha$, to wit
\begin{equation}\label{6.2}
    \Psi = \Psi^{(0)} + \alpha\, \Psi^{(1)}+ \ldots\, .
\end{equation}
The problem we are facing now is that for non-zero $\alpha$ the constraints are no longer first class and therefore we need a nonstandard procedure to handle them. One possibility would be Gupta--Bleuler quantization \cite{KowalskiGlikman:1992pn}, but the required procedure of splitting the constraints into holomorphic and anti-holomorphic parts is technically complex and, presumably, leads to explicit breaking of Lorentz covariance (see \cite{Sengupta:2009kg} for discussion in a similar context.) Another possibility would be to make use of the master constraint program \cite{Klauder:1996nx}, \cite{Thiemann:2003zv}, \cite{Dittrich:2004bp}, and
\cite{Sengupta:2009kg}, but this is again technically involved.

Instead we adopt the definition of physical wave function $\Psi$ such that the matrix elements of all the constraints are zero
\begin{equation}\label{6.3}
    \left<\Psi\right| \Phi \left|\Psi\right> =0\, ,
\end{equation}
which is a weakened version of Gupta--Bleuler scheme. It should be stressed that the expression (\ref{6.3}) is formal, because to make the precise sense of it we must specify the inner product in the Hilbert space of states.

 Now we use (\ref{6.3}) to define the perturbative theory in $\alpha$. In the zeroth order we have
\begin{equation}\label{6.4}
    \left<\Psi^{(0)}\right| \Phi^{(0)} \left|\Psi^{(0)}\right> =0\, ,
\end{equation}
while in the first order in $\alpha$ we find
$$
    \left<\Psi^{(1)}\right| \Phi^{(0)} \left|\Psi^{(0)}\right> +\left<\Psi^{(0)}\right| \Phi^{(0)} \left|\Psi^{(1)}\right> +$$\begin{equation}\label{6.5}\left<\Psi^{(0)}\right| \Phi^{(1)} \left|\Psi^{(0)}\right> =0\, .
\end{equation}
Inspecting (\ref{r1x})--(\ref{r1d}) we find that it follows from (\ref{6.4}), (\ref{6.5}) that the zeroth order wave function has to satisfy the following four conditions
\begin{eqnarray}
          0 &=&  \left<\Psi^{(0)}\right| i\frac{\delta}{\delta e_i^\alpha(x)} \left|\Psi^{(0)}\right>   \\
         0 &=& \left<\Psi^{(0)}\right| i\frac{\delta}{\delta \omega_i^{\alpha\beta}(x)} \left|\Psi^{(0)}\right>    \\
          0 &=&  \left<\Psi^{(0)}\right| \epsilon^{ijk}K_{\alpha\beta}{}^{\gamma\delta}\, \cD^\omega_i \Big(e_{j\,\gamma}e_{k\,\delta}\Big) \left|\Psi^{(0)}\right> \\
           0 &=& \left<\Psi^{(0)}\right|\epsilon^{ijk}\,K_{\alpha\beta}{}^{\gamma\delta}\, e^{\,\,\beta}_{i}\,F_{jk\,\,\gamma\delta} \left|\Psi^{(0)}\right> \end{eqnarray}
Knowing $\Psi^{(0)}$ one can turn to the remaining first order equation, resulting from (\ref{r1y}), along with some of the second order ones, to find $\Psi^{(1)}$, and then go to the next order analysis. We stop the discussion at this point leaving the details to another paper.

\section{Conclusions}

In this paper we have performed the canonical analysis of the constrained $\SO(4,1)$ BF theory. This analysis, although quite involved, seems to be significantly simpler than the analogous one of Plebanski theory reported in \cite{Buffenoir:2004vx}, leading however to the slightly more general effective description of the dynamical degrees of freedom provided by Holst constraints that include Immirzi parameter. This suggests that it might be not only simpler, but also more natural to consider spin foam model associated with this particular formulation of gravity. Unfortunately, not much work has been done till now on the $\SO(4,1)$ spin foam models, which would require to handle somehow not only the quadratic $B$ field term, but also the representation theory of $\SO(4,1)$ group, which is more complicated than the one of $\SU(2)$ group, usually used in the spin foam context.

\begin{acknowledgments}
We would like to thank Alejandro Perez and Michal Szczachor for helpful discussion. For JKG this work is supported in part by Research Project No.~N202 081 32/1844 and No.~NN202318534 and by Polish Ministry of Science and Higher Education Grant 182/N-QGG/2008/0.
\end{acknowledgments}


\begin{thebibliography}{99.}

%\cite{Ashtekar:1986yd}
\bibitem{Ashtekar:1986yd}
  A.~Ashtekar,
  ``New Variables for Classical and Quantum Gravity,''
  Phys.\ Rev.\ Lett.\  {\bf 57} (1986) 2244.
  %%CITATION = PRLTA,57,2244;%%

  %\cite{Rovelli:2004tv}
\bibitem{Rovelli:2004tv}
  C.~Rovelli,
  ``Quantum Gravity,''
%\href{http://www.slac.stanford.edu/spires/find/hep/www?irn=5994683}{SPIRES entry}
{\it  Cambridge, UK: Univ. Pr. (2004) 455 p}

%\cite{Thiemann:2007zz}
\bibitem{Thiemann:2007zz}
  T.~Thiemann,
  ``Modern canonical quantum general relativity,''
%\href{http://www.slac.stanford.edu/spires/find/hep/www?irn=7656084}{SPIRES entry}
{\it  Cambridge, UK: Cambridge Univ. Pr. (2007)}



  %\cite{Barbero:1994ap}
\bibitem{Barbero:1994ap}
  J.~F.~Barbero G.,
  ``Real Ashtekar variables for Lorentzian signature space times,''
  Phys.\ Rev.\  D {\bf 51} (1995) 5507
  [arXiv:gr-qc/9410014].
  %%CITATION = PHRVA,D51,5507;%%

  %\cite{Immirzi:1996di}
\bibitem{Immirzi:1996di}
  G.~Immirzi,
 ``Real and complex connections for canonical gravity,''
  Class.\ Quant.\ Grav.\  {\bf 14} (1997) L177
  [arXiv:gr-qc/9612030].
  %%CITATION = CQGRD,14,L177;%%

  %\cite{Rovelli:1996dv}

%\cite{Holst:1995pc}
\bibitem{Holst:1995pc}
  S.~Holst,
  ``Barbero's Hamiltonian derived from a generalized Hilbert-Palatini action,''
  Phys.\ Rev.\  D {\bf 53} (1996) 5966
  [arXiv:gr-qc/9511026].
  %%CITATION = PHRVA,D53,5966;%%

\bibitem{Rovelli:1996dv}
  C.~Rovelli,
  ``Black hole entropy from loop quantum gravity,''
  Phys.\ Rev.\ Lett.\  {\bf 77} (1996) 3288
  [arXiv:gr-qc/9603063].
  %%CITATION = PRLTA,77,3288;%%


  %\cite{Rezende:2009sv}
\bibitem{Rezende:2009sv}
  D.~J.~Rezende and A.~Perez,
  ``4d Lorentzian Holst action with topological terms,''
  Phys.\ Rev.\  D {\bf 79}, 064026 (2009)
  [arXiv:0902.3416 [gr-qc]].
  %%CITATION = PHRVA,D79,064026;%%

  %\cite{MacDowell:1977jt}
\bibitem{MacDowell:1977jt}
  S.~W.~MacDowell and F.~Mansouri,
  ``Unified Geometric Theory Of Gravity And Supergravity,''
  Phys.\ Rev.\ Lett.\  {\bf 38} (1977) 739
  [Erratum-ibid.\  {\bf 38} (1977) 1376].
  %%CITATION = PRLTA,38,739;%%

    %\cite{Plebanski:1977zz}
\bibitem{Plebanski:1977zz}
  J.~F.~Plebanski,
  ``On the separation of Einsteinian substructures,''
  J.\ Math.\ Phys.\  {\bf 18}, 2511 (1977).
  %%CITATION = JMAPA,18,2511;%%


  %\cite{Freidel:2005ak}
\bibitem{Freidel:2005ak}
  L.~Freidel and A.~Starodubtsev,
  ``Quantum gravity in terms of topological observables,''
  arXiv:hep-th/0501191.
  %%CITATION = HEP-TH/0501191;%%

  %\cite{Smolin:2003qu}
\bibitem{Smolin:2003qu}
  L.~Smolin and A.~Starodubtsev,
  ``General relativity with a topological phase: An action principle,''
  arXiv:hep-th/0311163.
  %%CITATION = HEP-TH/0311163;%%


%\cite{Randono:2008bb}
\bibitem{Randono:2008bb}
  A.~Randono,
  ``A New Perspective on Covariant Canonical Gravity,''
  Class.\ Quant.\ Grav.\  {\bf 25}, 235017 (2008)
  [arXiv:0805.3169 [gr-qc]].
  %%CITATION = CQGRD,25,235017;%%

  %\cite{Date:2008rb}
\bibitem{Date:2008rb}
  G.~Date, R.~K.~Kaul and S.~Sengupta,
  ``Topological Interpretation of Barbero-Immirzi Parameter,''
  Phys.\ Rev.\  D {\bf 79} (2009) 044008
  [arXiv:0811.4496 [gr-qc]].
  %%CITATION = PHRVA,D79,044008;%%


 % \cite{GarciaCompean:1999kj}
\bibitem{GarciaCompean:1999kj}
  H.~Garcia-Compean, O.~Obregon, C.~Ramirez and M.~Sabido,
  ``Remarks on 2+1 self-dual Chern-Simons gravity,''
  Phys.\ Rev.\  D {\bf 61}, 085022 (2000)
  [arXiv:hep-th/9906154].
  %%CITATION = PHRVA,D61,085022;%%

  %\cite{Mercuri:2007ki}
\bibitem{Mercuri:2007ki}
  S.~Mercuri,
  ``From the Einstein-Cartan to the Ashtekar-Barbero canonical constraints,
  passing through the Nieh-Yan functional,''
  Phys.\ Rev.\  D {\bf 77} (2008) 024036
  [arXiv:0708.0037 [gr-qc]].
  %%CITATION = PHRVA,D77,024036;%%

  %\cite{Kodama:1990sc}
\bibitem{Kodama:1990sc}
  H.~Kodama,
  ``Holomorphic Wave Function Of The Universe,''
  Phys.\ Rev.\  D {\bf 42} (1990) 2548.
  %%CITATION = PHRVA,D42,2548;%%


%\cite{Freidel:2003pu}
\bibitem{Freidel:2003pu}
  L.~Freidel and L.~Smolin,
  ``The linearization of the Kodama state,''
  Class.\ Quant.\ Grav.\  {\bf 21}, 3831 (2004)
  [arXiv:hep-th/0310224].
  %%CITATION = CQGRD,21,3831;%%

  %\cite{KowalskiGlikman:1992pn}
\bibitem{KowalskiGlikman:1992pn}
  J.~Kowalski-Glikman,
  ``On The Gupta-Bleuler Quantization Of The Hamiltonian Systems With
  Anomalies,''
  Annals Phys.\  {\bf 232} (1994) 1
  [arXiv:hep-th/9211028].
  %%CITATION = APNYA,232,1;%%



  %\cite{Sengupta:2009kg}
\bibitem{Sengupta:2009kg}
  S.~Sengupta,
  ``Quantum realizations of Hilbert-Palatini second-class constraints,''
  arXiv:0911.0593 [gr-qc].
  %%CITATION = ARXIV:0911.0593;%%

  %\cite{Klauder:1996nx}
\bibitem{Klauder:1996nx}
  J.~R.~Klauder,
  ``Coherent state quantization of constraint systems,''
  Annals Phys.\  {\bf 254} (1997) 419
  [arXiv:quant-ph/9604033].
  %%CITATION = APNYA,254,419;%%

  %\cite{Thiemann:2003zv}
\bibitem{Thiemann:2003zv}
  T.~Thiemann,
  ``The Phoenix project: Master constraint programme for loop quantum
  gravity,''
  Class.\ Quant.\ Grav.\  {\bf 23} (2006) 2211
  [arXiv:gr-qc/0305080].
  %%CITATION = CQGRD,23,2211;%%

%\cite{Dittrich:2004bp}
\bibitem{Dittrich:2004bp}
  B.~Dittrich and T.~Thiemann,
  ``Testing the master constraint programme for loop quantum gravity. II:
  Finite dimensional systems,''
  Class.\ Quant.\ Grav.\  {\bf 23} (2006) 1067
  [arXiv:gr-qc/0411139].
  %%CITATION = CQGRD,23,1067;%%

  %\cite{Buffenoir:2004vx}
\bibitem{Buffenoir:2004vx}
  E.~Buffenoir, M.~Henneaux, K.~Noui and Ph.~Roche,
  ``Hamiltonian analysis of Plebanski theory,''
  Class.\ Quant.\ Grav.\  {\bf 21} (2004) 5203
  [arXiv:gr-qc/0404041].
  %%CITATION = CQGRD,21,5203;%%




\end{thebibliography}
\end{document}